\documentclass[12pt]{article}
\usepackage{ascmac}
\usepackage{ulem}
\usepackage{latexsym}
\usepackage{graphicx}
\usepackage{amsmath}
\usepackage[top=3.5cm,bottom=3.5cm,left=2cm,right=2cm]{geometry}
\usepackage{pifont}
\usepackage{cancel}             
\usepackage{amsmath,amssymb}

\newcommand{\h}{\hspace}
\newcommand{\aln}[1]{\begin{align}#1\end{align}}

\newcommand{\be}{\begin{equation}}
\newcommand{\e}{\end{equation}}
\usepackage{color}

\begin{document}

\title{
\vbox{
\baselineskip 14pt
\hfill \hbox{\normalsize KUNS-2577
}} \vskip 1cm
\bf \Large Natural solution to the naturalness problem\\
$-$ Universe does fine-tuning $-$
\vskip 0.5cm
}
\author{
Yuta~Hamada\thanks{E-mail:  \tt hamada@gauge.scphys.kyoto-u.ac.jp}, 
Hikaru~Kawai\thanks{E-mail:  \tt hkawai@gauge.scphys.kyoto-u.ac.jp} and
Kiyoharu~Kawana\thanks{E-mail: \tt kiyokawa@gauge.scphys.kyoto-u.ac.jp} 
\bigskip\\
\it \normalsize
 Department of Physics, Kyoto University, Kyoto 606-8502, Japan\\
\smallskip
}
\date{\today}

\maketitle

\abstract{\normalsize We propose a new mechanism to solve the fine-tuning problem. We start from a multi-local action $ S=\sum_{i}c_{i}S_{i}+\sum_{i,j}c_{i,j}S_{i}S_{j}+\sum_{i,j,k}c_{i,j,k}S_{i}S_{j}S_{k}+\cdots$, where $S_{i}$'s are ordinary local actions. Then, the partition function of this system is given by 
\be Z=\int d\overrightarrow{\lambda} f(\overrightarrow{\lambda})\langle f|T\exp\left(-i\int_{0}^{+\infty}dt\hat{H}(\overrightarrow{\lambda};a_{cl}(t))\right)|i\rangle,\nonumber\e
where $\overrightarrow{\lambda}$ represents the parameters of the system whose Hamiltonian is given by $\hat{H}(\overrightarrow{\lambda};a_{cl}( t))$, $a_{cl}(t)$ is the radius of the universe determined by the Friedman equation, and $f(\overrightarrow{\lambda})$, which is determined by $S$, is a smooth function of $\overrightarrow{\lambda}$. If a value of $\overrightarrow{\lambda}$, $\overrightarrow{\lambda}_{0}$, dominates in the integral, we can interpret that the parameters are dynamically tuned to $\overrightarrow{\lambda}_{0}$. We show that indeed it happens in some realistic systems. In particular, we consider the strong CP problem, multiple point criticality principle and cosmological constant problem. It is interesting that these different phenomena can be explained by one mechanism.

\newpage
\section{Introduction and general idea}\label{sec:int}
Since the discovery of the Higgs particle \cite{Aad:2012tfa,Chatrchyan:2012ufa}, it has become more and more important to consider the fine-tuning problem \cite{Weinberg:1988cp,Martin:2012bt,Baker:2006ts}. The natural and conservative approach is to seek solutions in the context of ordinary local field theory. However, even if such theory can solve an individual problem such as the quadratic divergence of the Higgs mass, it seems difficult to answer the whole problems simultaneously. Therefore, it is necessary to consider a new framework or principle beyond ordinary local field theory. There are many interesting proposals such as the asymptotic safety \cite{Shaposhnikov:2009pv}, hidden duality and symmetry \cite{Kawamura:2013kua,Kawamura:2013xwa,Kawamura:2015ffa},  classical conformality \cite{Meissner:2006zh,Meissner:2007xv,Iso:2009ss,Iso:2009nw,Iso:2012jn}, multiple point criticality principle (MPP) \cite{Froggatt:1995rt,Froggatt:2001pa,Nielsen:2012pu,Buttazzo:2013uya,Hamada:2012bp,Bezrukov:2007ep,Hamada:2013mya,Hamada:2014iga,Hamada:2014wna,Hamada:2014raa,Bezrukov:2014bra,Haba:2014sia,Hamada:2015ria,Hamada:2014xka,Kawana:2014zxa,Kawana:2015tka,Haba:2015rha,Hamada:2015fma} and maximum entropy principle \cite{Kawai:2011qb,Kawai:2013wwa,Hamada:2014ofa,Hamada:2014xra,Hamada:2015wea}.

Among them, Coleman's theory on baby universe and multiverse \cite{Coleman:1988tj} seems promising. Although Coleman's first work explains the smallness of the cosmological constant (CC), its validity is unclear because it is based on Euclidean gravity. Therefore, its Lorentzian version is inevitably needed to give reliable predictions. In \cite{Kawai:2011qb,Kawai:2013wwa,Hamada:2014ofa,Hamada:2014xra}, such improvements are actually done. 
They are essentially based on the multi-local action and the existence of multiverse. The former is relatively natural because it arises if we take the topology change into account \cite{Coleman:1988tj,Coleman:1988cy,Kawana:2014vra,Asano:2012mn}. On the other hand, the idea of multiverse is slightly uncommon because there is no evidence so far that we live in multiverse. Therefore, it is meaningful to consider whether we can solve the fine-tuning problem without relying on the existence of multiverse. The purpose of this paper is to give a new framework to solve the fine-tuning problem based on the multi-local action.

We consider the partition function of the multi-local action:  
\be S_{M}=\sum_{i}c_{i}S_{i}+\sum_{i,j}c_{i,j}S_{i}S_{j}+\sum_{i,j,k}c_{i,j,k}S_{i}S_{j}S_{k}+\cdots,\label{eq:multiac}\e
where
\be S_{i}=\int_{0}^{\infty}dt\int d^{3}x{\cal{O}}_{i}(t,\mbox{\boldmath $x$})\e
is an ordinary local action, and $c_{i}$'s, $c_{i,j}$'s, $\cdots$ are constants. Here, we have assumed that the universe is created at $t=0$, and evolves to $t=\infty$. Then, Eq.(\ref{eq:multiac}) is obtained after summing up the wormholes.  
See Fig.\ref{fig:multi} for example. Thus, by expressing $\exp\left(iS_{M}\right)$ as a Fourier transform 
\be \exp\left(iS_{M}\right)=\int d\overrightarrow{\lambda} f(\overrightarrow{\lambda})\exp\left(i\sum_{i}\lambda_{i}S_{i}\right),\e
we can write the partition function of the system as follows:
\begin{align} Z&=\int_{t=0}^{t=\infty}{\cal{D}}\phi\exp\left(iS_{M}\right)\psi_{f}^{*}\psi_{i}\nonumber\\
&=\int d\overrightarrow{\lambda} f(\overrightarrow{\lambda})\int_{t=0}^{t=\infty}{\cal{D}}\phi\exp\left(i\sum_{i}\lambda_{i}S_{i}\right)\psi_{f}^{*}\psi_{i}\nonumber\\
&=
\int d\overrightarrow{\lambda} f(\overrightarrow{\lambda})\langle f|T\exp\left(-i\int_{0}^{+\infty}dt\hat{H}(\overrightarrow{\lambda};a_{cl}(t))\right)|i\rangle,\label{eq:partition}\end{align}
where $a_{cl}(t)$ is the radius of the universe determined by the Friedman equation, $\psi_{i}$ and $\psi_{f}$ represent initial and final states which are independent of $\overrightarrow{\lambda}$, and we have assumed that the universe eternally expands like our universe\footnote{Here, for simplicity, we have assumed that $a_{cl}(t)$ is already given. See Section \ref{sec:ccp} for more concrete argument.}. If there is a point $\overrightarrow{\lambda_{0}}$ that strongly dominates in the integral of Eq.(\ref{eq:partition}), the observer in the universe finds that the parameters are fixed at $\overrightarrow{\lambda}_{0}$. Namely, Eq.(\ref{eq:partition}) is equivalent to 
\be Z\sim
f(\overrightarrow{\lambda_{0}})\langle f|T\exp\left(-i\int_{0}^{+\infty}dt\hat{H}(\overrightarrow{\lambda_{0}};a_{cl}(t))\right)|i\rangle.\label{eq:partition1}\e
At a first glance, it seems difficult to evaluate Z because it involves the total history of the universe. However, in almost all the time, the universe is sufficiently  expanded, and its energy density is very close to that of the vacuum. More concretely, we have
\be T\exp\left(-i\int_{0}^{+\infty}dt\hat{H}(\overrightarrow{\lambda};a_{cl}(t))\right)|i\rangle\sim\exp\left(-i\varepsilon(\overrightarrow{\lambda})\int_{t^{*}}^{+\infty}dtV_{3}(a_{cl}(t))\right)|\psi(t^{*};\overrightarrow{\lambda})\rangle,\label{eq:vacuum}\e
where 
\be |\psi(t;\overrightarrow{\lambda})\rangle=T\exp\left(-i\int_{0}^{t}dt'\hat{H}(\overrightarrow{\lambda};a_{cl}(t'))\right)|i\rangle,\e
$V_{3}(a_{cl}(t))$ is the space volume, $\varepsilon(\overrightarrow{\lambda})$ is the vacuum energy density, and $t^*$ is a time such that the energy density of the state $|\psi(t;\overrightarrow{\lambda})\rangle$ is sufficiently closed to $\varepsilon(\overrightarrow{\lambda})$. By substituting Eq.(\ref{eq:vacuum}) to Eq.(\ref{eq:partition}), we obtain
\be Z\sim\int d\overrightarrow{\lambda} f(\overrightarrow{\lambda})\exp\left(-i\varepsilon(\overrightarrow{\lambda})\int_{t^{*}}^{+\infty}dtV_{3}(a_{cl}(t))\right)\langle f|\psi(t^{*};\overrightarrow{\lambda})\rangle.\label{eq:partition2}\e
As we will see in the following discussion, we can find $\overrightarrow{\lambda_{0}}$ from Eq.(\ref{eq:partition2}) in some phenomenologically interesting systems.

\begin{figure}
\begin{center}
\includegraphics[width=11cm]{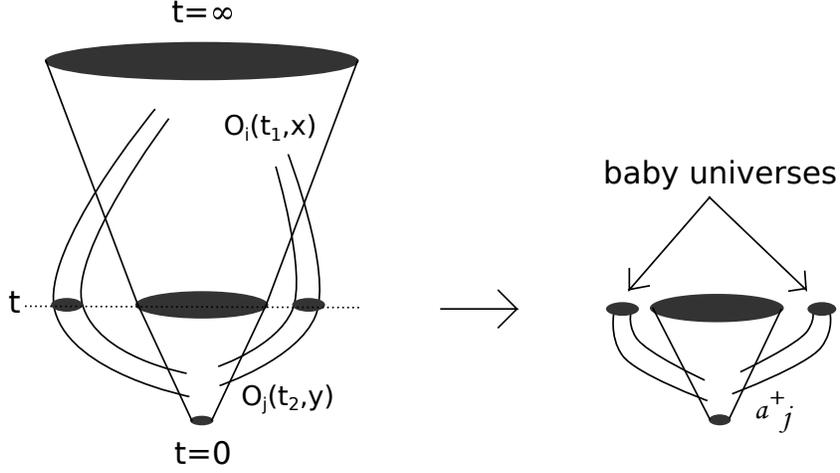}
\end{center}
\caption{Universe and baby universes. 
Here, we show wormholes having two legs. The left and right figures represent the partition function and Hamiltonian point of views respectively.  
}
\label{fig:multi}
\end{figure}

We can also repeat the above argument from the point of view of the Coleman's baby universe as follows. As discussed in \cite{Coleman:1988tj,Coleman:1988cy}, instead of using the multi-local action, the wormhole effect can be expressed by introducing operators $\hat{a}_{i}$ and $\hat{a}_{i}^{\dagger}$ that describe the creation and annihilation of a baby universe. Namely, the action of the universe with baby universes is given by
\be S_{C}=\sum_{i}(\hat{a}_{i}+\hat{a}_{i}^{\dagger})S_{i}.\label{eq:babyaction}\e
See the right figure in Fig.\ref{fig:multi} for example.  Because $\hat{a}_{i}+\hat{a}^{\dagger}_{i}$'s are conserved quantities, for each set of their eigenvalues, Eq.(\ref{eq:babyaction}) is an ordinary local action for the universe. Therefore, we can develop the quantum theory for the total system consisting of the universe and baby universes: 
\be \hat{H}_{\text{tot}}=\int d\overrightarrow{\lambda}|\overrightarrow{\lambda}\rangle\langle \overrightarrow{\lambda}|\otimes \hat{H}(\overrightarrow{\lambda};a_{cl}(t)).\e
Here, $\{|\overrightarrow{\lambda}\rangle\}$ is the complete set of the Fock space of the baby universes: 
\be 1=\int d\overrightarrow{\lambda}|\overrightarrow{\lambda}\rangle \langle\overrightarrow{\lambda}|\h{3mm},\h{3mm}\left(\hat{a}_{i}+\hat{a}_{i}^{\dagger}\right)|\overrightarrow{\lambda}\rangle=\lambda_{i}|\overrightarrow{\lambda}\rangle,\e
and $\hat{H}(\overrightarrow{\lambda};a_{cl}(t))$ is the Hamiltonian for the universe that corresponds to the action $\sum_{i}\lambda_{i}S_{i}$.
Then, for the initial state $|i\rangle =|i\rangle_{\text{baby}}\otimes|i\rangle_{\text{universe}}$, the wave function at $t$ is given by
\aln{ e^{-i\hat{H}_{\text{tot}}t}|i\rangle&=\int d\overrightarrow{\lambda}|\overrightarrow{\lambda}\rangle\langle \overrightarrow{\lambda}|i\rangle_{\text{baby}} \otimes T\exp\left(-i\int_{0}^{t}dt'\hat{H}(\overrightarrow{\lambda};a_{cl}(t'))\right)|i\rangle_{\text{universe}}
.\label{eq:finitetime}}
Thus, by considering the $t\rightarrow+\infty$ limit, and multiplying a final state $\langle f|:={}_{\text{baby}}\langle f|\otimes {}_{\text{universe}}\langle f|$ to Eq.(\ref{eq:finitetime}), we actually obtain Eq.(\ref{eq:partition}) where 
\be f(\overrightarrow{\lambda})={}_{\text{baby}}\langle f|\overrightarrow{\lambda}\rangle\langle\overrightarrow{\lambda}|i\rangle_{\text{baby}}.\e
This is the derivation of Eq.(\ref{eq:partition}) from the point of view of the Coleman's baby universe. Note that the weight function $f(\overrightarrow{\lambda})$ is not important in the following discussion because we will consider the general consequences which do not depend on the detail of $f(\overrightarrow{\lambda})$ as long as it is a smooth function. In the following, we do not consider such a special case that $|i\rangle_{\text{baby}}$ or $|f\rangle_{\text{baby}}$ is $|\overrightarrow{\lambda}'\rangle$. 

Because the above point of view is rather uncommon, let us see a simple example here: 
\begin{align} S&=\int_{-\infty}^{+\infty}dt\h{1mm}\left(\frac{m}{2}\dot{x}^{2}-\frac{m\omega_{0}^{2}}{2}x^{2}\right)+\frac{m^{2}\kappa}{4^{2}}\int_{-\infty}^{+\infty}dt\h{1mm}x^{2}\int_{-\infty}^{+\infty}dt\h{1mm}x^{2}\nonumber\\
&:=S_{0}+\kappa S_{H}^{2},
\end{align}
where 
\be S_{0}=\int_{-\infty}^{+\infty}dt\h{1mm}\left(\frac{m}{2}\dot{x}^{2}-\frac{m\omega_{0}^{2}}{2}x^{2}\right)\h{2mm},\h{2mm}S_{H}=\frac{m}{4}\int_{-\infty}^{+\infty}dt \h{1mm} x^{2}.\e
This is a harmonic oscillator having an additional bi-local action $S_{H}^{2}$. By using a Lagrangian multiplier $\lambda_{1}$, the path integral of this system can be rewritten as follows:
\begin{align} Z&=\int{\cal{D}}x \exp\left(iS_{0}+i\kappa S_{H}^{2}\right)\nonumber\\ 
&= \sqrt{\frac{-i\kappa}{\pi}}\int d\lambda_{1} \h{1mm}e^{-i\kappa\lambda_{1}^{2}}\int{\cal{D}}x \exp\left(iS_{0}+2i\kappa\lambda_{1} S_{H}\right)\nonumber\\
&:=\int d\lambda f(\lambda)\int{\cal{D}}x\h{1mm}e^{i\tilde{S}(\lambda)}
,\label{eq:harmpart}
\end{align}
where 
\be \tilde{S}(\lambda)=\int_{-\infty}^{+\infty} dt\h{1mm}\left(\frac{m}{2}\dot{x}^{2}-\frac{m\lambda }{2}x^{2}\right)\h{2mm},\h{2mm} f(\lambda):=\sqrt{\frac{-i}{\kappa\pi}}\times e^{-i\frac{(\omega_{0}^{2}-\lambda)^{2}}{\kappa}}.\e
Here, in the last line in Eq.(\ref{eq:harmpart}), we have changed the variable as
\be\lambda:=\omega_{0}^{2}-\kappa\lambda_{1}.\e
One can see that Eq.(\ref{eq:harmpart}) actually has the form of Eq.(\ref{eq:partition}). 
However, $\lambda$ is not fixed to any special value because the integrand in Eq.(\ref{eq:harmpart}) is a regular function of $\lambda$ in this case.
On the other hand, as we will see, if we consider systems with infinite degrees of freedom, $\overrightarrow{\lambda}$ is indeed fixed to a special value in several cases. 
\\

In the following sections, we study a few examples which show that the above argument is not an armchair theory. In particular, we consider the strong CP problem, MPP and cosmological constant problem (CCP). 
The MPP is a proposal to solve the fine-tuning problem, which claims that, 
when a field theory has two vacua, the parameters are fixed so that they become degenerate. In regard to the CCP, we will see that the CC is nearly fixed to zero by generalizing Eq.(\ref{eq:partition}) to the Wheeler-Dewitt wave function. In order to explain the positive small CC, we will give some argument based on the existence of multiverse. 
\\

\noindent 
This paper is organized as follows. In Section \ref{sec:math}, we give a few important mathematical preliminaries. In Section \ref{sec:strong}, we discuss the strong CP problem. In Section \ref{sec:MPP}, we derive the MPP from Eq.(\ref{eq:partition}). In Section \ref{sec:ccp}, we consider the CCP by generalizing Eq.(\ref{eq:partition}) to the Wheeler-Dewitt wave function. 
In Section \ref{sec:summary}, we give summary and discussion.

\section{Mathematical preliminaries}\label{sec:math}
In this section, we consider the singular behavior of $e^{ikg(\lambda)}$ in the $k\rightarrow\infty$ limit, which plays a crucial role in the following discussion. 

If $g(\lambda)$ is smooth, and has a stationary point $\lambda_{0}$, we obtain
\be e^{ig(\lambda)k}\underset{k\rightarrow\infty}{\sim}\sqrt{\frac{2\pi}{ikg''(\lambda)}}e^{ikg(\lambda_{0})}\delta(\lambda-\lambda_{0}),\label{eq:saddl}\e
by using the saddle point approximation. Note that the right hand side is suppressed by the factor $\sqrt{k}$.

If $g(\lambda)$ is smooth and monotonic in the $\lambda>0$ region, and $g'(0)\neq0$, we have 
\be e^{ikg(\lambda)}\theta(\lambda)\underset{k\rightarrow\infty}{\sim}\frac{i}{k}\left(\frac{dg}{d\lambda}\right)^{-1}e^{ikg(0)}\delta(\lambda),\label{eq:delta}\e
where $\theta(\lambda)$ is a step function. 
The proof is as follows. By multiplying a test function $F(\lambda)$ with finite support to $e^{ikg(\lambda)}$, and integrating from 0 to $\infty$, we obtain
\aln{ \int_{0}^{\infty}d\lambda e^{ig(\lambda)k}F(\lambda)&=\int_{g(0)}^{\infty}dg\h{1mm}\left(\frac{dg}{d\lambda}\right)^{-1}e^{ikg}F(\lambda=\lambda(g))\nonumber\\
&=\left[\frac{e^{ikg}}{ik}\left(\frac{dg}{d\lambda}\right)^{-1}F(\lambda(g))\right]_{g(0)}^{\infty}+{\cal{O}}\left(\frac{1}{k^{2}}\right)\nonumber\\
&=\frac{i}{k}\left(\frac{dg}{d\lambda}\right)^{-1} e^{ikg(0)}F(\lambda)\Biggl|_{\lambda=0}+{\cal{O}}\left(\frac{1}{k^{2}}\right),\label{eq:delta1}}
Thus, one can see that Eq.(\ref{eq:delta}) holds in the $k\rightarrow\infty$ limit. Note that the right hand side is proportional to $1/k$, which is small compared with Eq.(\ref{eq:saddl}). 

Similarly, we can obtain the following equation for $g(\lambda)$ that is smooth and monotonic in each of the regions, $\lambda>\lambda_{0}$ and $\lambda<\lambda_{0}$: 
\be e^{ikg(\lambda)}\underset{k\rightarrow\infty}{\sim}\frac{i}{k}\left[e^{ikg(\lambda)}\left(\frac{dg}{d\lambda}\right)^{-1}\Biggl|_{\lambda_{0}+}-e^{ikg(\lambda)}\left(\frac{dg}{d\lambda}\right)^{-1}\Biggl|_{\lambda_{0}-}\right]\delta(\lambda-\lambda_{0}).\label{eq:delta2}\e
Note that the right hand side is non-zero only when $g(\lambda)$ is not smooth at $\lambda_{0}$. We will call such a point $\lambda_{0}$ non-analytic point.

\section{Strong CP problem}\label{sec:strong}
In the QCD Lagrangian, there exists a CP violating topological term 
\be S_{\theta}:=\frac{\theta}{16\pi^{2}}\int d^{4}xF^{a}_{\mu\nu}\tilde{F}^{a\mu\nu}.\e 
Experimentally, there is a strong upper bound on $\theta$ \cite{Baker:2006ts,Afach:2015sja}:
\be \theta<10^{-9},\e
which is unnaturally small. This is the so-called ``strong CP problem''. We can naturally solve this problem as follows.

From the general argument of Section \ref{sec:int}, the partition function is given by 
\aln{ Z&\sim\int_{0}^{2\pi} d\theta f(\theta)\exp\left(-i\varepsilon(\theta)\int_{t^{*}}^{+\infty}dtV_{3}(a_{cl}(t))\right)\langle f|\psi(t^{*};\theta)\rangle\nonumber\\
&=\int_{0}^{2\pi} d\theta f(\theta)\exp\left(-i\varepsilon(\theta)V_{4}\right)\langle f|\psi(t^{*};\theta)\rangle,\label{eq:qcd}}
where 
\be \varepsilon(\theta)\sim\Lambda_{\text{QCD}}^{4}\cos\theta\e
is the energy density of the $\theta$ vacuum (see \cite{Coleman:1978ae} for example), and $V_{4}$ is the volume of the space time. 
If we consider such a final state that has a finite winding number\footnote{If $|f\rangle$ is the $n$ vacuum $|n\rangle$ with a large value of $n\sim \Lambda_{\text{QCD}}^{4}V_{4}$, the exponent of the integrand in Eq.(\ref{eq:qcd}) becomes stationary at the point where $\sin\theta\sim n/\left(\Lambda_{\text{QCD}}^{4}V_{4}\right)$.
},
 using Eq.(\ref{eq:saddl}), we obtain \footnote{If $k$ is finite in Eq.(\ref{eq:saddl}), we have a function of width $1/\sqrt{kg''(\lambda)}$ instead of the delta function. Then, the width $\Delta\theta$ is given by $\sqrt{1/\left(V_{4}\Lambda_{QCD}^{4}\right)}\sim10^{-80}/\left(\sqrt{V_{4}H_{0}^{4}}\right)$ where $H_{0}$ the present Hubble constant. 
}
\be 
\exp\left(-i\varepsilon(\theta)V_{4}\right)\sim\sqrt{\frac{2\pi}{iV_{4}\Lambda_{\text{QCD}}^{4}}}\left(e^{-i\varepsilon(0)V_{4}}\delta(\theta)+e^{-i\varepsilon(\pi)V_{4}}\delta(\theta-\pi)\right).\e
By substituting this to Eq.(\ref{eq:qcd}), the partition function becomes 
\be { Z\sim\sqrt{\frac{1}{V_{4}\Lambda_{\text{QCD}}^{4}}}\Biggl[f(0)e^{-i\varepsilon(0)V_{4}}\langle f|\psi(t^{*};0)\rangle+f(\pi)e^{-i\varepsilon(\pi)V_{4}}\langle f|\psi(t^{*};\pi)\rangle\Biggl].
\label{eq:qcd1}}\e
This shows that the partition function is strongly dominated by $\theta=0$ and $\theta=\pi$ worlds, as in the many world interpretation.

\section{Multiple point criticality principle}\label{sec:MPP}
\begin{figure}
\begin{center}
\includegraphics[width=8.5cm]{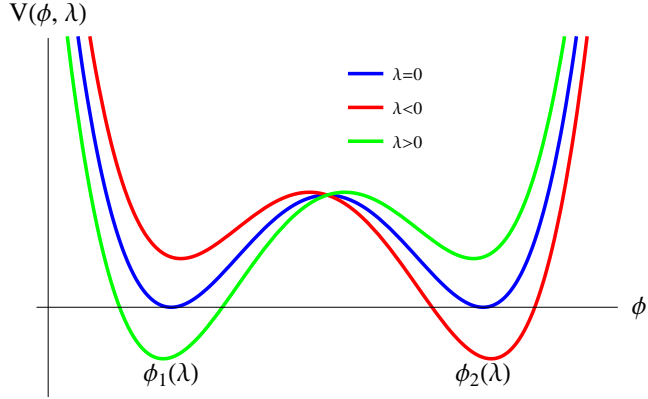}
\end{center}
\caption{Schematic behavior of $V(\phi,\lambda)$. The blue line corresponds to $\lambda=0$, and the green (red) line corresponds to $\lambda>0$ $(<0)$.}
\label{fig:MPP}
\end{figure}
In this section, we derive the MPP by assuming that the potential $V(\phi,\lambda)$ of a scalar field $\phi$ has two minima at $\phi_{1}(\lambda)$ and $\phi_{2}(\lambda)$, where we take $\phi_{1}(\lambda)<\phi_{2}(\lambda)$. Here, $\lambda$ is one of the coupling constants of the theory. We assume that two minima become degenerate when $\lambda$ is equal to zero
, and that the signature of $\lambda$ is chosen as 
\aln{V(\phi_{1}(\lambda),\lambda)&<V(\phi_{2}(\lambda),\lambda)\nonumber\text{ \h{5mm}for $\lambda>0$,}\\
\\
V(\phi_{1}(\lambda),\lambda)&>V(\phi_{2}(\lambda),\lambda)\nonumber\text{ \h{5mm}for $\lambda<0$}.}
See Fig.\ref{fig:MPP} for example. Then, the true vacuum expectation value $\phi_{\text{vac}}(\lambda)$ and the vacuum energy density $\varepsilon(\lambda)$ are given by 
\be \phi_{\text{vac}}(\lambda)=\begin{cases}\phi_{2}(\lambda)&\text{ for $\lambda<0$}\\
\phi_{1}(\lambda)&\text{ for $\lambda>0$}\end{cases}\h{3mm},\h{3mm}
\varepsilon(\lambda)=\begin{cases}V(\phi_{2}(\lambda))&\text{for $\lambda<0$}\\
V(\phi_{1}(\lambda))&\text{for $\lambda>0$}.\end{cases}\label{eq:vac2}\e
According to Eq.(\ref{eq:partition}), the partition function is given by 
\be Z=\int d\lambda f(\lambda)\exp\left(-i\varepsilon(\lambda)V_{4}\right)\langle f|\psi(t^{*};\lambda)\rangle.\label{eq:timeMPP}\e
We assume that $V(\phi_{1}(\lambda))$ and $V(\phi_{2}(\lambda))$ are monotonic functions of $\lambda$ in $\lambda>0$ and $\lambda<0$ respectively, and that their derivatives are not equal at $\lambda=0$. Because for generic $|f\rangle$, $\langle f| \psi(t^{*};\lambda)\rangle$ has no singularity at $ \lambda=0$, we can use Eq.(\ref{eq:delta2}):
\be 
e^{-i\varepsilon(\lambda)V_{4}}\sim-\frac{ie^{-i\varepsilon(0)V_{4}}}{V_{4}}\times\left[\left(\frac{V(\phi_{1}(\lambda))}{d\lambda}\right)^{-1}\Biggl|_{0+}-\left(\frac{V(\phi_{2}(\lambda))}{d\lambda}\right)^{-1}\Biggl|_{0-}\right]\delta(\lambda).\e
By substituting this into Eq.(\ref{eq:timeMPP}), we obtain 
\be Z\sim\frac{f(0)}{V_{4}}\times e^{-i\varepsilon(0)V_{4}} \langle f|\psi(t^{*});0\rangle.\e
Thus we have derived the MPP in the context of the multi-local action. 

\section{Generalization to Wheeler-Dewitt wave function}\label{sec:ccp}
In this section, we consider the generalization of Eq.(\ref{eq:partition}) to the Wheeler-Dewitt wave function and the fine tuning of the physical CC $\Lambda$. Unfortunately, it is not easy to consider the negative region $\Lambda<0$ because we do not know what happens after the Big Crunch. On the other hand, in the positive region $\Lambda>0$, we can consider the Wheeler-Dewitt wave function for the entire region of the radius of the universe, and examine which value of $\overrightarrow{\lambda}$ dominates. Therefore, in the following, we take only the region $\Lambda>0$ into account in the path integral. 


Then the generalization of Eq.(\ref{eq:partition}) to the Wheeler-Dewitt wave function is given by 
\be Z_{WD}=\int d\Lambda_{B} \int d\overrightarrow{\lambda} f(\Lambda_{B},\overrightarrow{\lambda})\theta(\Lambda)\int_{0}^{\infty}dT\h{1mm}\langle f_{a}|\otimes\langle f_{MR}|e^{-i\left(\hat{H}_{G}(\Lambda_{B})+\hat{H}_{MR}(\overrightarrow{\lambda};\hat{a})\right)T}|\epsilon\rangle\otimes|i_{MR}\rangle,\label{eq:wave0}\e
where $\hat{H}_{G}(\Lambda_{B}):=-\hat{p}_{a}^{2}/\left(2\hat{a}M_{pl}^{2}\right)+\Lambda_{B}$ is the Hamiltonian for the radius of the universe $a$ with the  bare CC $\Lambda_{B}$  \cite{Hamada:2014ofa}, $\hat{H}_{MR}(\overrightarrow{\lambda};\hat{a})$ represents the other degrees of freedom, that is, the matter and radiation including gravitons, and $|\epsilon\rangle\otimes|i_{MR}\rangle$ ($|f_{a}\rangle\otimes|f_{MR}\rangle$) is an initial (a final) state of the universe. (We have assumed that the initial universe has the radius $a=\epsilon$.) The integration over $T$ comes from the path integral for the lapse function. In the following discussion, we assume $|f_{a}\rangle=|a_{\infty}\rangle$ where $a_{\infty}$ represents the large radius of the universe. Before considering the CCP, we first rederive the results of the previous sections from  Eq.(\ref{eq:wave0}). 

\subsection{Fixing the other parameters than cosmological constant}\label{eq:unified}
Eq.(\ref{eq:wave0}) differs from Eq.(\ref{eq:partition}) in that it contains the integration over the time $T$. However, the results in the previous sections can be 
also obtained from Eq.(\ref{eq:wave0}) because, for a fixed value of $a$, the $T$ integral is dominated by $T_{a}$ at which the radius of the universe becomes $a$. 
More concretely, we have
\aln{ \int_{0}^{\infty}dT\h{1mm}\langle a|\otimes\langle f_{MR}|e^{-i\left(\hat{H}_{G}(\Lambda_{B})+\hat{H}_{MR}(\overrightarrow{\lambda};\hat{a})\right)T}|\epsilon\rangle\otimes |i_{MR}\rangle&\sim\langle f_{MR}|T\exp\left(-i\int_{0}^{T_{a}}dt\hat{H}_{MR}(\overrightarrow{\lambda};a_{cl}(t))\right)|i_{MR}\rangle\nonumber\\
&:=\langle f_{MR}|\psi_{MR}(T_{a})\rangle,\label{eq:wave1}}
where we have omitted the integrations over $\Lambda_{B}$ and $\overrightarrow{\lambda}$ for simplicity. 
Here, $a_{cl}(t)$ satisfies the following Friedman equation and the boundary condition:
\aln{ H^{2}:=\left(\frac{\dot{a_{cl}}}{a_{cl}}\right)^{2}&=\frac{1}{3M_{pl}^{2}}\left(\Lambda_{B}+\frac{\langle \psi_{MR}(t)|\hat{H}_{MR}(\overrightarrow{\lambda};a_{cl}(t))|\psi_{MR}(t)\rangle}{V_{3}(a_{cl}(t))}\right)\h{2mm},\h{2mm}a_{cl}(0)=\epsilon,\label{eq:born0}}
where $V_{3}(a_{cl}(t))$ is the volume of the space. 
We can understand Eq.(\ref{eq:wave1}) within the Born-Oppenheimer approximation \cite{Born} by assuming that the expansion rate $H$ is very small compared with the energy scale of the other degrees of freedom. See \ref{app:time} for the details. 

Using Eq.(\ref{eq:wave1}), we can rederive the results of the previous sections. For a sufficiently large value of $a$, we can replace the Hamiltonian $\hat{H}_{MR}(\overrightarrow{\lambda};a_{cl}(T_{a}))$ by the vacuum energy $E_{0}(\overrightarrow{\lambda};a_{cl}(T_{a}))=\varepsilon(\overrightarrow{\lambda})V_{3}(a_{cl}(T_{a}))$ as in Eq.(\ref{eq:partition2}). 
Therefore, the right hand side of Eq.(\ref{eq:wave1}) becomes  
\aln{ \exp\left(-i\int_{0}^{T_{a}}dt\hat{H}_{MR}(\overrightarrow{\lambda},a_{cl}(t))\right)|i_{MR}\rangle&\sim\exp\left(-i\varepsilon(\overrightarrow{\lambda})\int_{t^*}^{T_{a}}dt V_{3}(a_{cl}(t))\right)|\psi_{MR}(t^*)\rangle
,}
where $t^*$ is a time such that the energy density of the state $|\psi_{MR}(t^*)\rangle$ is sufficiently closed to that of the vacuum. Then, Eq.(\ref{eq:wave0}) can be written as 
\aln{ \int d\Lambda_{B} \int d\overrightarrow{\lambda} f(\Lambda_{B},\overrightarrow{\lambda})\theta(\Lambda)\exp\left(-i\varepsilon(\overrightarrow{\lambda})\int_{t^*}^{T_{a}}dt V_{3}(a_{cl}(t))\right)\langle f_{MR}|\psi_{MR}(t^*)\rangle.}
Thus, by repeating the same argument, we can rederive the results of the previous sections.

\subsection{Solving the cosmological constant problem}
In order to fix $\Lambda$, it is sufficient to consider the effective action for the radius $a$ because only the vacuum energy is relevant:  
\be Z=\int_{0}^{\infty}dT\int_{0}^{\infty}d\Lambda f(\Lambda)\langle a_{\infty}|e^{-i\hat{H}(\Lambda)T}|\epsilon\rangle,\label{eq:wave4} \e
where the effective Hamiltonian $\hat{H}(\Lambda)$ is
\be \hat{H}(\Lambda)=-\frac{\hat{p}_{a}^{2}}{2\hat{a}M_{pl}^{2}}+\frac{\hat{a}^{3}\rho(\hat{a})}{6}\h{2mm},\h{2mm}\rho(\hat{a})=\Lambda+\rho_{MR}(\hat{a}).\label{eq:uniham}\e
Here, $\rho_{MR}(\hat{a})$ stands for the energy density of the matter and radiation including gravitons. One can see that $-\hat{a}^{4}\rho(\hat{a})$ plays the roll of a potential of the radius of the universe. By inserting the complete set 
\be 1=\int_{-\infty}^{+\infty} dE|E;\Lambda\rangle\langle E;\Lambda|\h{2mm},\h{2mm}\hat{H}(\Lambda)|E;\Lambda\rangle=E|E;\Lambda\rangle\e
into Eq.(\ref{eq:wave4}),  we obtain
\aln{ Z&=\int_{0}^{\infty} d\Lambda f(\Lambda)\int_{0}^{\infty}dt\int_{-\infty}^{+\infty}dE\h{1mm}e^{-iEt}\langle a_{\infty}|E;\Lambda\rangle\langle E;\Lambda|\epsilon\rangle\nonumber\\
&=\int_{0}^{\infty} d\Lambda f(\Lambda)\left(\pi\langle a_{\infty}|0;\Lambda\rangle\langle 0;\Lambda|\epsilon\rangle+PV\int_{-\infty}^{+\infty}\frac{dE}{E}\langle a_{\infty}|E;\Lambda\rangle\langle E;\Lambda|\epsilon\rangle\right),\label{eq:wave2}}
where $|0;\Lambda\rangle$ is the zero-energy eigenstate, and $PV$ represents the principal value. Here, we have used the following identity \footnote{The derivation is as follows: We can neglect $e^{-iEt}$ by introducing the adiabatic factor $E\rightarrow E-i\epsilon$. Thus, by multiplying a smooth test function $F(E)$ with finite support to the left hand side, and integrating over $E$, we have
\be \int_{-\infty}^{\infty}dE\frac{-1}{-i\left(E-i\epsilon\right)}F(E)=\pi F(i\epsilon)-PV\int_{-\infty}^{\infty}dE\frac{F(E)}{E-i\epsilon}.\nonumber\e 
Therefore, in the $\epsilon\rightarrow0$ limit, we obtain Eq.(\ref{eq:pv}). Here, note that, if we also include the negative region in the time integral, we obtain the delta function $2\pi\delta(E)$ instead of Eq.(\ref{eq:pv}). This leads to the ordinary Wheeler-Dewitt state $|0;\Lambda\rangle$.}:
\be \lim_{t\rightarrow+\infty}\frac{e^{-iEt}-1}{-iE}=\pi\delta(E)+PV\frac{1}{E}.\label{eq:pv}\e 
The second term in Eq.(\ref{eq:wave2}) comes from the fact that we have chosen $t=0$ as the beginning of the universe. However, because the universe is well described classically by the Friedman equation, we expect that only the small region around $E=0$ is relevant in this integral. In \ref{app:Eint}, we actually check this by using the WKB approximation. As a result, Eq.(\ref{eq:wave2}) can be approximated by
\be Z\sim\int_{0}^{\infty} d\Lambda f(\Lambda)\langle a_{\infty}|0;\Lambda\rangle\langle 0;\Lambda|\epsilon\rangle.\label{eq:wave3}\e
Let us now evaluate $\langle a|0;\Lambda\rangle$ by using the WKB approximation. The Wheeler-Dewitt wave function is given by \cite{Hamada:2014ofa}
\be \langle a|0;\Lambda\rangle=M_{pl}\sqrt{\frac{a}{p_{cl}}}\exp\left(i\int^{a} da'\h{1mm}p_{cl}(a')\right),\h{2mm}p_{cl}(a):=M_{pl}a^{2}\sqrt{\frac{\rho(a)}{3}},\label{eq:wkb1}\e
where $p_{cl}(a)$ is the classical momentum. 
For simplicity, we consider the matter dominated universe:
\be \rho(a)=\Lambda+\frac{M}{a^{3}}:=\Lambda+\rho_{M}(a),\e
where $M$ is the total energy of the matter. Then, we can evaluate the exponent in Eq.(\ref{eq:wkb1}):
\aln{\int_{a_{M}}^{a} da'\h{1mm}p_{cl}(a')&=\frac{M_{pl}}{3^{\frac{3}{2}}}\left(a^{3}\sqrt{\rho(a)}-a_{M}^{3}\sqrt{\rho(a_{M})}+\frac{M}{\sqrt{\Lambda}}\log\left[\frac{a^{\frac{3}{2}}(\Lambda+\sqrt{\Lambda\rho(a)})}{a_{M}^{\frac{3}{2}}(\Lambda+\sqrt{\Lambda\rho(a_{M})})}\right]\right)\nonumber\\
&:=\frac{M_{pl}a^{3}}{3^{\frac{3}{2}}}g(\Lambda,a),\label{eq:wkbexponent}}
where $a_{M}$ is the radius of the universe at the time when the matter dominated era starts. $g(\Lambda,a)$ is a smooth and monotonic function of $\Lambda$, which satisfies 
\aln{ g(0,a)
=&2\left(\sqrt{\rho_{M}(a)}-\left(\frac{a_{M}}{a}\right)^{3}\sqrt{\rho_{M}(a_{M})}\right)\h{2mm},\h{2mm}\lim_{a\rightarrow\infty}g(\Lambda,a)=\sqrt{\Lambda},\nonumber\\
&\frac{dg(\Lambda,a)}{d\Lambda}\Biggl|_{\Lambda=0}=\frac{1}{3}\left(\frac{1}{\sqrt{\rho_{M}(a)}}-\left(\frac{a_{M}}{a}\right)^{3}\frac{1}{\sqrt{\rho_{M}(a_{M})}}\right).}
Thus, by substituting Eq.(\ref{eq:wkbexponent}) to Eq.(\ref{eq:wkb1}), and using Eq.(\ref{eq:delta}), we obtain
\aln{ \langle a|0;\Lambda\rangle&\underset{a\rightarrow\infty}{\sim}\frac{3^{\frac{3}{2}}i}{M_{pl}a^{3}}\left(\frac{dg(\Lambda,a)}{d\Lambda}\right)^{-1}\Biggl|_{\Lambda=0}M_{pl}\sqrt{\frac{a}{p_{cl}(a)}}\exp\left(i\frac{M_{pl}a^{3}}{3^{\frac{3}{2}}}g(0,a)\right)\delta(\Lambda)\nonumber\\
&=\frac{3^{\frac{3}{2}}i}{M_{pl}a^{3}}\left(\frac{dg(\Lambda,a)}{d\Lambda}\right)^{-1}\Biggl|_{\Lambda=0}\delta(\Lambda)\langle a|0;0\rangle,}
where $|0;0\rangle$ is the zero energy eigenstate with $\Lambda=0$. By substituting this to Eq.(\ref{eq:wave3}), we have
\aln{Z&\sim\frac{1}{a_{\infty}^{3}}\left(\frac{dg(\Lambda,a_{\infty})}{d\Lambda}\right)^{-1}\Biggl|_{\Lambda=0}\langle a_{\infty}|0;0\rangle\langle0;0|\epsilon\rangle\nonumber\\
&\sim\frac{1}{a_{\infty}^{3}}\left(\frac{dg(\Lambda,a_{\infty})}{d\Lambda}\right)^{-1}\Biggl|_{\Lambda=0}\int_{0}^{\infty}dt\langle a_{\infty}|e^{-i\hat{H}(0)t}|\epsilon\rangle.\label{eq:effwhee}}
This indicates that the whole universe is described by the Wheeler-Dewitt wave function with $\Lambda=0$. However, this result is inconsistent with the cosmological observation \cite{Planck:2015xua}, which supports the small but non-zero $\Lambda$. In the next section, we discuss a possibility to explain the discrepancy by assuming the multiverse.

\section{Summary and discussion}\label{sec:summary}
We have proposed a new mechanism to solve the fine-tuning problem of the universe: Assuming the multi-local action, we have obtained the partition function Eq.(\ref{eq:partition}) or its generalization to the Wheeler-Dewitt wave function Eq.(\ref{eq:wave0}). We have found that, in some phenomenologically interesting cases, there is a special point $\overrightarrow{\lambda}_{0}$ that strongly dominates in the partition function. This fact can be understood as the dynamical fine-tuning of the parameters. 
In particular, we have solved the strong CP problem and CCP, and also derived the MPP. To obtain these results, we have assumed that 
the final state of the universe $|f\rangle_{\text{universe}}$ is fixed to a generic state. Although the justification of this assumption remains an open problem, it is interesting and remarkable that the different phenomena of field theory are explained by an unique mechanism.

Finally,  let us discuss a possibility to obtain the small but nonzero CC. As we will see in the following, we can obtain the fluctuation of the CC that is consistent with the observed value if we assume (i) the existence of the multiverse, and (ii) that only the region within the horizon is relevant to examine the partition function. Taking the wormhole effects into consideration \cite{Kawai:2011qb,Kawai:2013wwa,Hamada:2014ofa,Hamada:2014xra}, we obtain the generalized partition function $Z_{M}$ of the multiverse instead of Eq.(\ref{eq:partition}):
\aln{Z_{M}:&=
\sum_{N=0}^{\infty}\int dg\frac{f(g)}{N!}Z_{U}(g)^{N}
=\int dgf(g)\exp\left(Z_{U}(g)\right),\label{eq:multiparti}}
where $g$ is a coupling constant, and $Z_{U}(g)$ is the partition function of a single universe. We have assumed that all the universes are copies of our universe. 
Eq.(\ref{eq:multiparti}) indicates a new possibility to fix the parameter $g$: It is fixed to the saddle point $g^{*}$ of $Z_{U}(g)$ even if $Z_{U}(g)$ itself does not have a strong peak as $\delta(g-g^{*})$. $Z_{U}(g)$ can be expanded as 
\be Z_{U}(g)=Z_{U}(g^{*})+\frac{1}{2}\frac{d^{2}Z_{U}}{dg^{2}}\Biggl|_{g=g^{*}}(g-g^{*})^{2}+{\cal{O}}\left((g-g^{*})^{3}\right).\e
From the path integral expression,
\be Z_{U}(g)=
\int {\cal{D}}\phi e^{igS_{g}+\cdots}\times\psi_{f}^{*}\psi_{i},\e
$d^{2}Z_{U}/dg^{2}|_{g=g^{*}}$ can be expressed as the expectation value of the local action $S_{g}$:
\aln{ \frac{d^{2}Z_{U}}{dg^{2}}\Biggl|_{g=g^{*}}&=-\int {\cal{D}}\phi e^{ig^{*}S_{g^{*}}+\cdots}S_{g^{*}}^{2}\times \psi_{f}^{*}\psi_{i}\nonumber\\
&:=-\langle \hat{S}_{g^{*}}^{2}\rangle Z_{U}(g^{*}).
}
Therefore, the fluctuation of $g$ around $g^{*}$ is given by
\be \Delta g\sim\left(\frac{d^{2}Z_{U}}{dg^{2}}\right)^{-\frac{1}{2}}\Biggl|_{g=g^{*}}=\frac{1}{\sqrt{\langle\hat{S}_{g^{*}}^{2}\rangle Z_{U}(g^{*})}}.\e
Typically, $\langle\hat{S}_{g^{*}}^{2}\rangle\propto TV_{3}$ because
\aln{ \langle\hat{S}_{g^{*}}^{2}\rangle&=\int d^{4}x\int d^{4}y\langle \underbrace{\hat{{\cal{O}}}_{g^{*}}(x)\hat{{\cal{O}}}_{g^{*}}(y)}_{\text{contract}}\rangle\nonumber\\
&=\int d^{4}x\int d^{4}y W(x-y)\nonumber\\
&=V_{4}\int d^{4}xW(x)\sim V_{4}M_{pl}^{4},}
where we have assumed that the cut-off scale is $M_{pl}$, and that
$\int d^{4}xW(x)\sim M_{pl}^{4}$
by the dimensional analysis. Thus, one can see that $\Delta g$ is of order $(V_{4}M_{pl}^{4}Z_{U}(g^{*}))^{-\frac{1}{2}}$. This leads to the fluctuation of the vacuum energy density $\Delta \rho_{0}$, which is typically
\be \Delta \rho_{0}\sim M_{pl}^{4}\Delta g\sim\frac{M_{pl}^{4}}{\sqrt{V_{4}M_{pl}^{4}Z_{U}(g^{*})}}\sim \frac{M_{pl}^{2}H_{0}^{2}}{\sqrt{Z_{U}(g^{*})}},\e
where we have replaced $V_{4}$ with $H_{0}^{-4}$. Thus, if $Z_{U}(g^{*})$ is ${\cal{O}}(1)$, this fluctuation is consistent with the observed value. Although this conclusion is based on a few nontrivial assumptions, it is interesting that such a small CC can be obtained by maximizing the partition function as a function of the parameter.

In conclusion, the theory of the multi-local action is attractive and promising to solve the fine-tuning problem of the universe.

\section*{Acknowledgement} 
This work is supported by the Grant-in-Aid for Japan Society for the Promotion of Science (JSPS) Fellows 
No.25$\cdot$1107 (YH) and No.27$\cdot$1771 (KK).

\appendix 
\def\thesection{Appendix \Alph{section}}
\section{Born-Oppenheimer approximation}\label{app:time}
In this Appendix, we give a justification of Eq.(\ref{eq:wave1}) based on the Born-Oppenheimer approximation. Because the time scale of $a$ is much longer than that of the matter and radiation, the total wave function can be obtained as following. 
\\
{\it{\textbf{Step1}}}: We first solve the Schr\"{o}dinger equation for the matter and radiation assuming that $a_{cl}(t)$ is a slowly changing c-number function of $t$: 
\be -i\frac{\partial}{\partial t}|\psi_{MR}(t)\rangle=\hat{H}_{MR}(a_{cl}(t))|\psi_{MR}(t)\rangle.\e
Then, the expectation value of the energy as function of $a$ is given
\be E_{MR}(a,t):=\langle \psi_{MR}(t)|\hat{H}_{MR}(a)|\psi_{MR}(t)\rangle.\label{eq:energy1}\e
\\
{\it{\textbf{Step2}}}: By using $E_{MR}(a,t)$, we solve the Schr\"{o}dinger equation for $a$:
\be -i\frac{\partial}{\partial t}|\psi_{r}(t)\rangle=\left(\hat{H}_{G}+E_{MR}(\hat{a})\right)|\psi_{r}(t)\rangle.\e
Then, we identify $a_{cl}(t)$ with the expectation value of the radius
\be a_{cl}(t)=\langle \psi_{r}(t)|\hat{a}|\psi_{r}(t)\rangle.\e
By solving Eq.(63)-(66) in a self-consistent manner, we obtain 
\aln{ \langle a|\otimes\langle f_{MR}&|e^{-i\left(\hat{H}_{G}(\Lambda_{B})+\hat{H}_{MR}(\overrightarrow{\lambda};\hat{a})\right)t}|\epsilon\rangle\otimes |i_{MR}\rangle \nonumber\\
&\sim\langle a|\psi_{r}(t)\rangle \langle f_{MR}|T\left\{\exp\left(-i\int_{0}^{t}dt'\hat{H}_{MR}(a_{cl}(t'))\right)\right\}|i_{MR}\rangle.\label{eq:born1}}
Then, within the Born-Oppenheimer approximation, the Wheeler-Dewitt wave function is given by
\be \langle a|\otimes\langle f_{MR}||\psi\rangle=\int_{0}^{\infty}dt \uwave{\langle a|\psi_{r}(t)\rangle} \langle f_{MR}|T\left\{\exp\left(-i\int_{0}^{t}dt'\hat{H}_{MR}(a_{cl}(t'))\right)\right\}|i_{MR}\rangle.\label{eq:born2}\e
Here, because $\langle a|\psi_{r}(t)\rangle$ has a peak at $a_{cl}(t)$ as a function of $a$, the $t$ integral of Eq.(\ref{eq:born2}) is dominated by $T_{a}$ such that   
\be a_{cl}(T_{a})\sim a.\e
Thus, we obtain Eq.(\ref{eq:wave1}).


\section{Evaluation of the Principal value}\label{app:Eint}
In this Appendix, we evaluate 
\be PV\int_{-\infty}^{+\infty}\frac{dE}{E}\langle a|E;\Lambda\rangle\langle E;\Lambda|\epsilon\rangle\label{eq:apppv}\e
by using the WKB approximation. 
The WKB solution of $\langle a|E;\Lambda\rangle$ is given by 
\be \langle a|E;\Lambda\rangle=M_{pl}\sqrt{\frac{a}{p_{cl}}}\exp\left(i\int^{a} da'p_{cl}(a')\right),\label{eq:wkb2}\e
where
\be p_{cl}(a)=M_{pl}a^{2}\sqrt{2\left(\frac{\rho(a)}{6}-\frac{E}{a^{3}}\right)}:=M_{pl}a^{2}\sqrt{\frac{\tilde{\rho}(a)}{3}}.\e
Then, for a sufficiently large value of $a$, we have
\aln{\int_{a_{M}}^{a} da\h{1mm}p_{cl}(a)
&=\frac{M_{pl}}{3^{\frac{3}{2}}}\left(a^{3}\sqrt{\tilde{\rho}(a)}-a_{M}^{3}\sqrt{\tilde{\rho}(a_{M})}+\frac{M-E}{\sqrt{\Lambda}}\log\left[\frac{a^{\frac{3}{2}}(\Lambda+\sqrt{\Lambda\tilde{\rho}(a)})}{a_{M}^{\frac{3}{2}}(\Lambda+\sqrt{\Lambda\tilde{\rho}(a_{M})})}\right]\right)\nonumber\\
&:=\frac{M_{pl}a^{3}}{3^{\frac{3}{2}}}g(\Lambda,E,a)
\underset{a\gg a_{M}}{\simeq}\frac{M_{pl}a^{3}}{3^{\frac{3}{2}}}\sqrt{\tilde{\rho}(a)},\label{eq:wkbexponent1}}
By substituting Eq.(\ref{eq:wkb2}) and Eq.(\ref{eq:wkbexponent1}) to Eq.(\ref{eq:apppv}), we obtain
\aln{ PV\int_{-\infty}^{\infty}\frac{dE}{E}\h{1mm}M_{pl}\sqrt{\frac{a}{p_{cl}}}\exp\left(i\frac{M_{pl}a^{3}}{3^{\frac{3}{2}}}\sqrt{\tilde{\rho}(a)}\right)\langle E;\Lambda|\epsilon\rangle}
By expanding the exponent around $E=0$, we have 
\be M_{pl}\sqrt{\frac{a}{p_{cl}}}\exp\left(i\frac{M_{pl}a^{3}}{3^{\frac{3}{2}}}\sqrt{\rho(a)}-i\frac{M_{pl}E}{3^{\frac{3}{2}}\sqrt{\rho(a)}}+{\cal{O}}(E^{2})\right)=\langle a|0;\Lambda\rangle\exp\left(-i\frac{M_{pl}E}{3^{\frac{3}{2}}\sqrt{\rho(a)}}+{\cal{O}}(E^{2})\right).\label{eq:E0expand}\e
Therefore, only the region 
\be |E|\lesssim\frac{\sqrt{\rho(a)}}{M_{pl}}\sim\frac{\sqrt{\Lambda}}{M_{pl}}\e
contributes to the integral \footnote{More concretely, by substituting Eq.(\ref{eq:E0expand}) to Eq.(\ref{eq:apppv}), and neglecting the $E$ dependence of $\langle E;\Lambda|\epsilon\rangle$, we obtain 
\be \langle a|0;\Lambda\rangle PV\int_{-\infty}^{a^{3}\rho(a)}\frac{dE}{E}\exp\left(-i\frac{M_{pl}E}{3^{\frac{3}{2}}\sqrt{\rho(a)}}\right)=\langle a|0;\Lambda\rangle PV\int_{-\infty}^{k}\frac{dx}{x}e^{-ix}\underset{k\rightarrow+\infty}{\rightarrow}\nonumber-i\pi\langle a|0;\Lambda\rangle,\e
where $k:=M_{pl}a^{3}\sqrt{\rho(a)}/3^{\frac{3}{2}}$.
 }. Therefore, it is self-consistent to show $\Lambda=0$ by using only the zero energy eigenstate $|0\rangle$.

\section{Symmetry enhancement}\label{app:ssr}
\begin{figure}[h!]
\begin{center}
\begin{tabular}{c}
\begin{minipage}{0.5\hsize}
\begin{center}
\includegraphics[width=7.5cm]{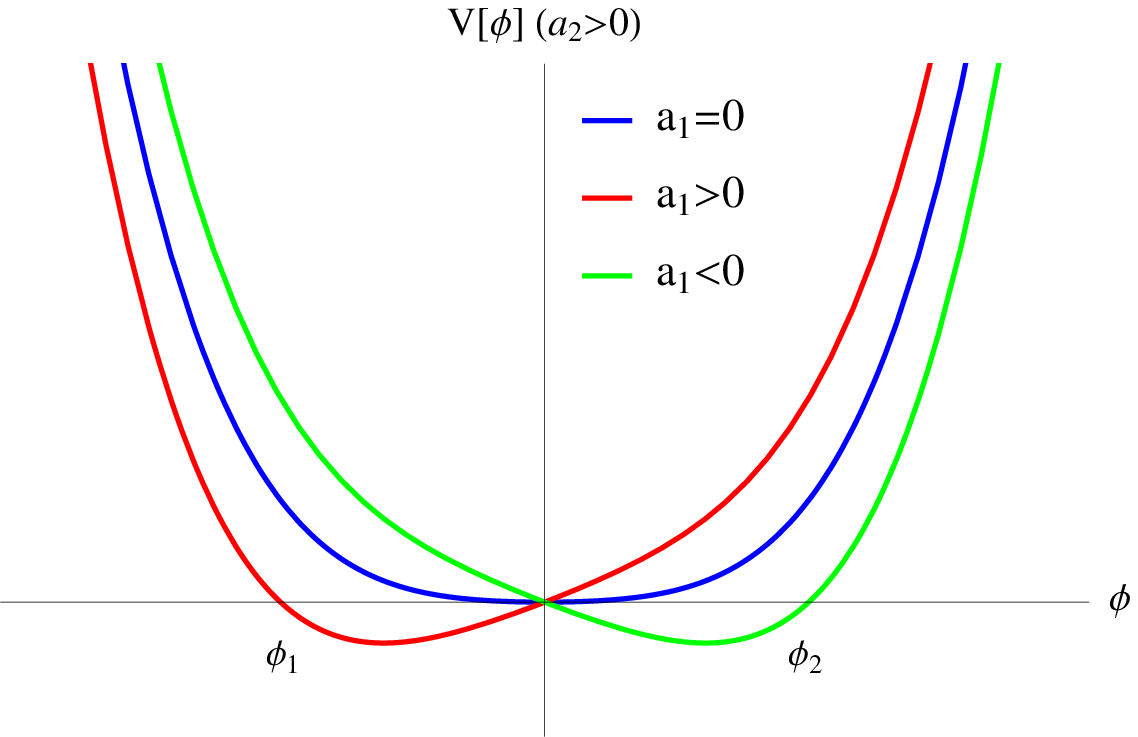}
\end{center}
\end{minipage}
\begin{minipage}{0.5\hsize}
\begin{center}
\includegraphics[width=7.5cm]{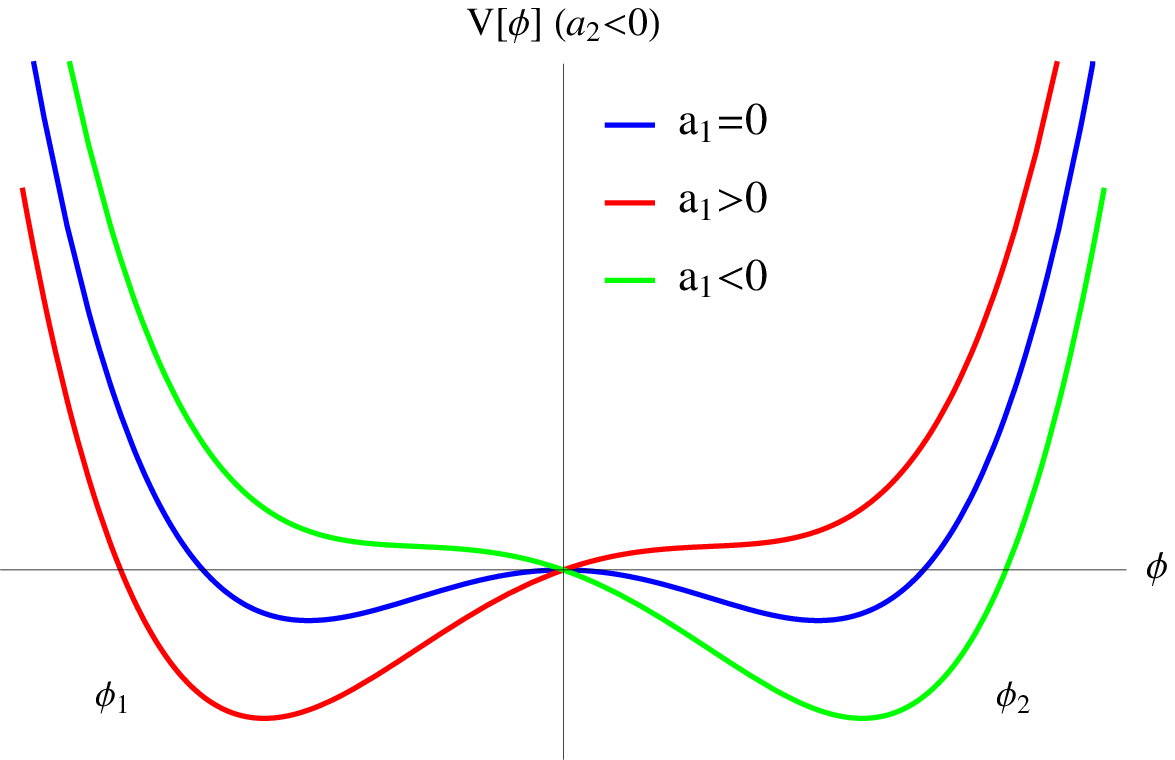}
\end{center}
\end{minipage}
\end{tabular}
\end{center}
\caption{Typical shapes of $V(\phi)$. The left (right) panel shows the $a_{2}>0$ $(<0)$ case.}
\label{fig:typical}
\end{figure}

In this Appendix, we study a mechanism by which symmetry is enhanced. In particular, we consider a scalar field $\phi$ with the effective potential 
\be V(\phi)=a_{1}\phi+\frac{a_{2}}{2}\phi^{2}+a_{3}\phi^{3}+\frac{a_{4}}{4}\phi^{4},
\e
where $a_{4}$ is fixed to a positive value so that the system is bounded below. We can eliminate $a_{3}\phi^{3}$ by shifting the field, $\phi\rightarrow\phi+\phi_{0}$, and  the potential becomes
\be V(\phi)=a_{1}\phi+\frac{a_{2}}{2}\phi^{2}+\frac{a_{4}}{4}\phi^{4}.\label{eq:pot1}\e
In Fig.\ref{fig:typical}, we show the typical shapes of $V(\phi)$. In the following discussion, we fix $a_{2}$ and vary $a_{1}$.  We denote the negative (positive) vacuum expectation value by $\phi_{1}(a_{1})$ $(\phi_{2}(a_{1}))$. According to the general argument in Section \ref{sec:int}, the partition function is 
\aln{ Z=
\int da_{1}f(a_{1})\exp\left(-i\varepsilon(a_{1})V_{4}\right)\langle f|\psi(t^{*};a_{ 1})\rangle
.\label{eq:sym1}}
where $\varepsilon(a_{1})$ is the vacuum energy density of this system, which is given by 
\be \varepsilon(a_{ 1})=\begin{cases}V(\phi_{1}(a_{1}))\h{5mm}&(\text{for $a_{1}>0$}),\\
\\
V(\phi_{2}(a_{1}))\h{5mm}&(\text{for $a_{1}<0$}).
\end{cases}\e
As a result, $a_{1}=0$ is apparently the non-analytic point of the vacuum energy $\varepsilon(a_{1})$. See Fig.\ref{fig:vac} for example. Thus, by using Eq.(\ref{eq:delta2}), we obtain
\be e^{-i\varepsilon(a_{1})V_{4}}\sim-\frac{ie^{-i\varepsilon(0)V_{4}}}{V_{4}}\times\left[\left(\frac{V(\phi_{1})}{da_{1}}\right)^{-1}\Biggl|_{a_{1}=0+}-\left(\frac{V(\phi_{2})}{da_{1}}\right)^{-1}\Biggl|_{a_{1}=0-}\right]\delta(a_{1}).\label{eq:a1fix}\e
By substituting this to Eq.(\ref{eq:sym1}), we have 
\be Z\sim\frac{1}{V_{4}}e^{-i\varepsilon(0)V_{4}}\langle f|\psi(t^{*};0)\rangle.\e
This indicates that the symmetric effective potential is favored by the multi-local action. 
\begin{figure}
\begin{center}
\includegraphics[width=9cm]{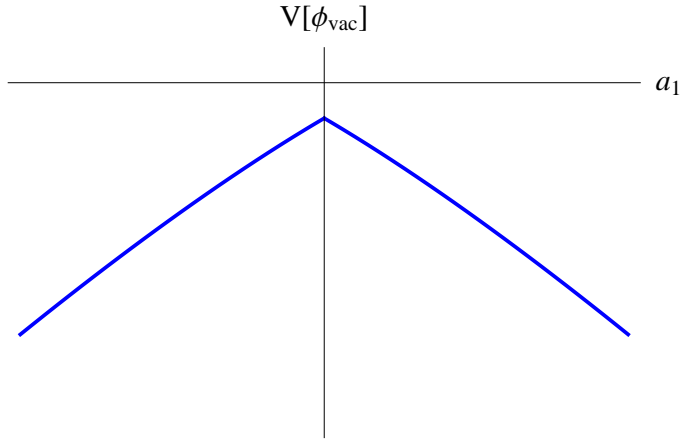}
\end{center}
\caption{The minimum of the potential as a function of $a_{1}$.}
\label{fig:vac}
\end{figure}

\end{document}